\documentclass[pdftex,preprint,twocolumn]{revtex4}
\usepackage[english]{babel}
\usepackage{times}
\usepackage{hyperref}
\usepackage[]{graphicx}

\begin{document}

\title{Average mass composition of primary
    cosmic rays in the superhigh energy region by Yakutsk complex EAS
    array data}

\author{S. P. Knurenko}\email[]{s.p.knurenko@ikfia.ysn.ru}
\author{A. A. Ivanov}\email[]{ivanov@ikfia.ysn.ru}
\author{A. V. Sabourov}\email[]{tema@ikfia.ysn.ru}
\author{I. Ye. Sleptsov}\email[]{i.ye.sleptsov.ikfia.ysn.ru}

\affiliation{Yu. G. Shafer Institute of Cosmophysical Research and
  Aeronomy, 31 Lenin Ave., 677980 Yakutsk, Russia}

\begin{abstract}
  The characteristics relating to the lateral and longitudinal
  development of EAS in the energy region of $10^{15} - 10^{19}$~eV
  have been analyzed in the framework of the QGSJET model and of mass
  composition of primary cosmic rays. It is found that at $E_{0} \ge 5
  \times 10^{15}$~eV the mean mass composition of primary cosmic rays
  begins to vary as indicated by a rise of $\left<\ln A\right>$ with
  increasing energy. The maximum value of $\left<\ln A\right>$ is
  observed at $E_{0} \sim (5 - 50) \times 10^{16}$~eV. It is confirmed
  by data of many compact EAS arrays and does not contradict an
  anomalous diffusion model of cosmic ray propagation in our
  Galaxy. In the superhigh energy region $(\ge 10^{18}$~eV) the value
  $\left<\ln A\right>$ begins to decrease, i.e. the mass composition
  becomes lighter and consists of protons and nuclei of He and C. It
  does not contradict our earlier estimations for the mass composition
  and points to a growing role of the metagalactic component of cosmic
  rays in the superhigh energy region.
\end{abstract}

\maketitle

\section{Introduction}

An investigation of superhigh energy cosmic rays in the 1950s - 1960s
and 1970s - 1990s has led to the discovery of irregularities in the
energy spectrum at the energy $\sim 3 \times 10^{15}$~eV and $\sim
10^{19}$~eV~\cite{1, 2}. The problem of origin of breaks in the energy
spectrum requires reliable knowledge about the mass composition of
primary cosmic rays (PCR) from high energies $\sim 10^{12} -
10^{14}$~eV up to ultra-high energies $10^{18} - 10^{21}$~eV.

The study of PCR mass composition at $E_{0} \ge 10^{15}$~eV is
possible only with the arrays registering extensive air showers (EAS)
and by indirect way only using the EAS characteristics maximum
sensitive to the mass composition. For these purposes, at the Yakutsk
complex array the characteristics of both longitudinal and radial
development of EAS are taken. The longitudinal development of showers
is reconstructed measuring the EAS Cherenkov light, and the radial
development measuring the structural functions of electron and muons
with $E_{\text{thr}} \ge 1$~GeV.

In this paper PCR mass composition data obtained at the Yakutsk EAS
array in recent year are summarized.

\section{Method}

The estimation of PCR mass composition was determined by the formula:
\begin{equation}
\left< \ln A\right> \equiv \sum_{i} a_{i} \cdot \ln A_{i} \text{,}
\label{eq1}
\end{equation}
where $a_{i}$ is the relative portion of nuclei to the mass number
$A_{i}$. In each case the experimental data were compared with
calculations by the QGSJET model for the primary proton and iron
nucleus in the framework of a superposition model:
\begin{equation}
\left<\ln A\right> = \frac{P^{\text{exp}} -
  P^{\text{p}}}{P^{\text{Fe}} - P^{\text{p}}} \times \ln A_{\text{Fe}}
\text{,}
\label{eq2}
\end{equation}
where $P$ is a parameter characterizing the longitudinal or radial
development of EAS, for example, $X_{\text{max}}$~--- a maximum depth
of a shower development, $\rho_{\mu}/\rho_{\text{s}}$~--- a portion of
muons at the fixed distance from a core, $N_{\mu} - N_{\text{s}}$~---
a correlation between the total number of muon and electrons at sea
level etc. The energy region of $10^{15} - 10^{19}$~eV, where there
are ``knee'' and  ``ankle'' irregularities in the CR energy spectrum,
has been considered specially.

\section{Results and discussion}

\begin{figure}
\centering
\includegraphics[width=0.45\textwidth, clip]{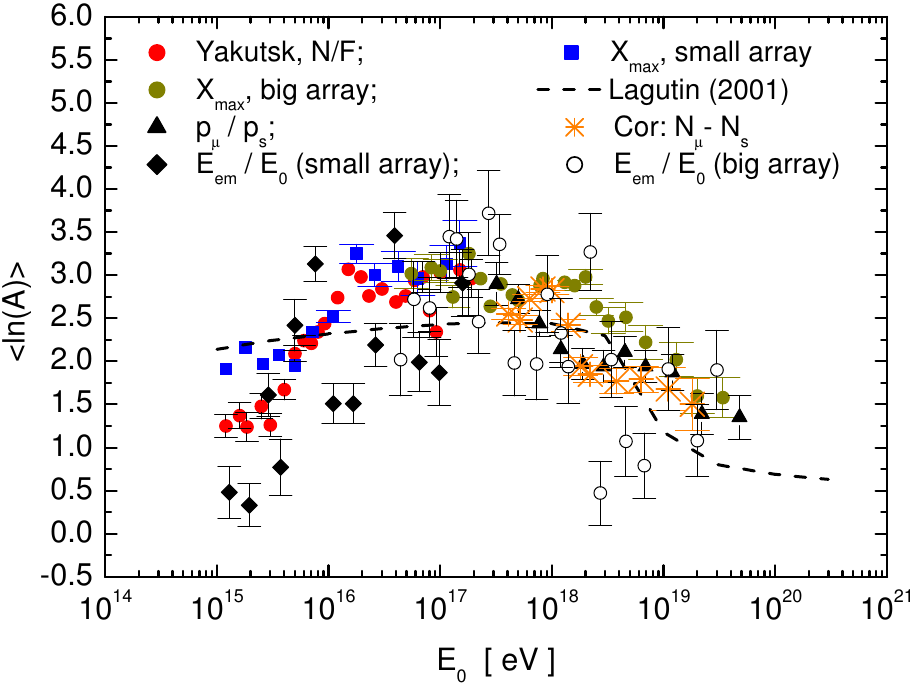}
\caption{Mass composition of CR obtained by different characteristics
  of EAS in the region of ultrahigh energies.}
\label{fig01}
\end{figure}

\begin{figure}
\centering
\includegraphics[width=0.45\textwidth, clip]{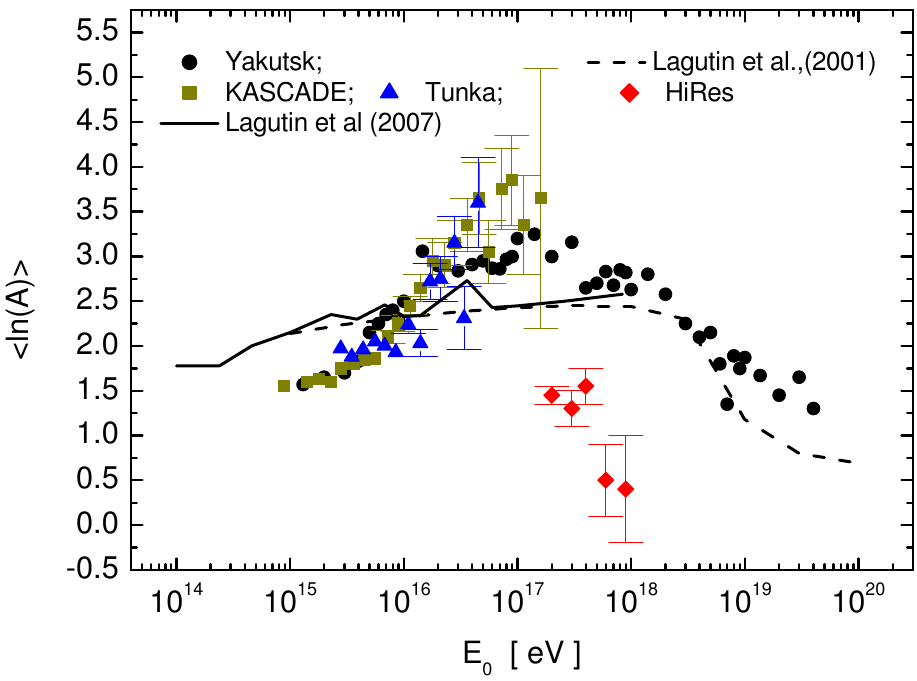}
\caption{Mass composition of cosmic rays at superhigh and ultrahigh
  energies. The curve is a calculation by Lagutin et al (2001) and
  (2007) according to the anomalous diffusion model for cosmic ray
  propogation.}
\label{fig02}
\end{figure}

Fig.~\ref{fig01} and Fig.~\ref{fig02} present results on the PCR mass
composition of the Yakutsk EAS array. Data were obtained in the
framework of the QGSJET model and two-component mass composition
(proton-iron nucleus). A few characteristics corresponding to the
radial and longitudinal development of EAS~\cite{3, 4, 5, 6} are used
in this analysis. In each case, the value $\left<\ln A\right>$ was
determined by an interpolation method according to~(\ref{eq2}). From
Fig.~\ref{fig01}, Fig.~\ref{fig02} it follows that the mass
composition changes reaching the heavier in the energy region of $(2 -
5) \times 10^{17}$~eV and then, beginning from $3 \times 10^{18}$~eV it
becomes lighter.

The curves are calculations according to the anomalous diffusion model
of cosmic ray propagation in the Galaxy (dashed line in
Fig.~\ref{fig01})~\cite{7} and the two-source model (solid line in
Fig.~\ref{fig02})~\cite{8}. The calculations have been carried out
separately for each of following groups of nuclei: p, He, CNO, N-Si,
Fe. The summary values of $\left<\ln A\right>$ are shown in
Fig.~\ref{fig01} and Fig.~\ref{fig02} and for the first case in the
Table~\ref{t01}. Data in the Table~\ref{t01} point to the change of
mass composition. In the energy region of $5 \times 10^{15} - 5 \times
10^{18}$~eV the mass composition is heavier than at $E_{0} \simeq
10^{19}$~eV. It contradicts the experimental data (Fig.~\ref{fig02}).
The calculations~\cite{8} according to a scenario for the particle
generation in the sources of two different types, assuming the
analogous model for the CR propagation, showed the presence of a fine
structure in the dependence of $\left<\ln A\right>$ on energy (see
Fig.~\ref{fig02}). There are also the sharp peaks in our experimental
data. In this connection, the investigation of the PCR mass
composition with using a great numbers of EAS characteristics and
statistical data as possible is of particular interest.

It should be noted that in the region after a ``knee'' the estimations
of the CR mass composition obtained at the compact arrays agree well
with each other. The same cannot be said of the energy region of $\sim
10^{18}$~eV (see Fig.~\ref{fig02}), where HiRes data point to the
faster enriching of the primary radiation by light nuclei and protons
in comparison with the Yakutsk array data, which point to the gradual
change of the mass composition from heavier to light (proton and He
nuclei, on the whole) at $E_{0} \sim 10^{19}$~eV. In both cases, the
data point to the existence of tendency to the ``protonization'' of
PCRs at $E_{0} >3 \times 10^{18}$~eV.

\begin{table*}
\centering
\caption{Calculation results for $\left<\ln A\right>$.}
\label{t01}
\renewcommand{\arraystretch}{1.3}
\begin{tabular}{ccccccccc}
\hline
 & A & H, \% &  He, \% & CNO, \% & Ni-Si, \% & Fe, \% & $\left<\ln
   A\right>$ & $\left<\ln A\right>$ \\
$E_{0}$, eV &  & & & & & & & experiment \\
\hline
$1 \times 10^{15}$ & & $22$ & $21$ & $20$ & $17$ & $19$ & $2.14$ &
$1.55 \pm 0.25$ \\
$1 \times 10^{16}$ & & $19$ & $19$ & $20$ & $18$ & $23$ & $2.32$ &
$2.15 \pm 0.24$ \\
$1 \times 10^{17}$ & & $17$ & $17$ & $20$ & $19$ & $26$ & $2.43$ &
$3.16 \pm 0.23$ \\
$1 \times 10^{18}$ & & $18$ & $16$ & $19$ & $19$ & $27$ & $2.44$ &
$2.61 \pm 0.21$ \\
$1 \times 10^{19}$ & & $55$ & $17$ & $9 $ & $ 8$ & $11$ & $1.18$ &
$1.66 \pm 0.18$ \\
$1 \times 10^{20}$ & & $68$ & $18$ & $ 6$ & $ 4$ & $ 3$ & $0.69$ &-\\
\hline
\end{tabular}
\end{table*}

\section{Summary}

A character of energy-dependence of $\left<\ln A\right>$ (see
Fig.~\ref{fig02}) by the Yakutsk array data points to the change of
primary particle mass composition at the singular points of  CR energy
spectrum. The $\left<\ln A\right>$ rises with energy after the
``knee'', riching the maximum value $\left<\ln A\right> = 3.5$ in the
  energy interval of $(2 - 5) \times 10^{17}$~eV, and then it begins
  to decrease. Such a behavior does not contradict a hypothesis for
  the propagation of CR by the anomalous diffusion laws in fractal
  interstellar medium (Lagutin et al, 2001) and according to which the
  mass composition of primary particles varies similar to experimental
  data. At $E_{0} > 10^{18}$~eV the value $\left<\ln A\right>$
  decreases gradually and at $E_{0} \sim 10^{19}$~eV the mass
  composition is practically represented by He nuclei and
  protons. This does not contradict calculations by Berezinsky et
  al.~\cite{9} for the metagalactic model, in which the ``ankle''
  observed in the experiments on registration of ultrahigh energy
  cosmic rays can only be formed by the proton component arriving from
  the Extragalaxy. Thereby, details of experimental spectrum form, for
  example, ``dip'', i.e. the decrease of intensity at $E_{0} \sim
  10^{19}$~eV, are most likely caused by the interaction of the
  extragalactic protons with relic photons $(\text{p} +
  \gamma_{\text{СМВ}} \to \text{p} + e^{-} + e^{+})$. As a direct
  evidence of this hypothesis, an anisotropy can be used which is
  associated with the origin and sources of cosmic rays. According
  to~\cite{10, 11, 12}, at $E_{0} > 8 \times 10^{18}$~eV the weak
  correlation in the arrival directions of EAS with the Galaxy plane
  and the strong correlation with the Extragalaxy plane are
  observed. The quasars can be possible sources of cosmic rays.

\end{document}